\documentclass[aps,preprint,nofootinbib,floatfix,superscriptaddress]{revtex4}
\usepackage{epsfig}
\usepackage{graphicx}
\usepackage{multirow}
\usepackage{amsmath}
\usepackage{epstopdf}
\usepackage{soul,xcolor}

\newcommand{\ktt}{|\bar 33\rangle}
\newcommand{\kss}{|6\bar 6\rangle}
\begin{document}
\setstcolor{blue}
\title{Exotic $bc\bar q\bar q$ four-quark states}
\author{T.~F.~Caram\'es}
\email{carames@usal.es}
\affiliation{Departamento de F{\'\i}sica Fundamental and IUFFyM,
Universidad de Salamanca, 37008 Salamanca, Spain}
\author{J.~Vijande}
\email{javier.vijande@uv.es}
\affiliation{Unidad Mixta de Investigaci\'on en Radiof\'\i sica e Instrumentaci\'on Nuclear 
en Medicina (IRIMED), Instituto de Investigaci\'on Sanitaria La Fe (IIS-La Fe)-Universitat de Valencia (UV) and 
IFIC (UV-CSIC), Valencia, Spain}
\author{A.~Valcarce}
\email{valcarce@usal.es}
\affiliation{Instituto Universitario de F\'\i sica Fundamental y Matem\'aticas (IUFFyM),
Universidad de Salamanca, 37008 Salamanca, Spain}
\date{\emph{Version of }\today}
\begin{abstract}
We carry out a systematic study of exotic $QQ^\prime \bar q\bar q$ four-quark states containing distinguishable 
heavy flavors, $b$ and $c$. Different generic constituent models are explored in an attempt to extract 
general conclusions. The results are robust, predicting the same sets of quantum numbers as the best candidates
to lodge bound states independently of the model used, the isoscalar $J^P=0^+$ and $J^P=1^+$ states. 
The first state would be strong and electromagnetic-interaction stable while the second
would decay electromagnetically to $\bar B D \gamma$. Isovector states are found to be unbound, 
preventing the existence of charged partners. The 
interest on exotic heavy-light tetraquarks with non identical heavy-flavors comes reinforced by
the recent estimation of the production rate of the isoscalar $bc\bar u \bar d$ $J^P=1^+$
state, two orders of magnitude larger than that of the $bb\bar u\bar d$ analogous state.
\end{abstract}
\maketitle
%%%
\section{Introduction}
\label{se:intro}
Among the flavor sectors where four-quark bound states may exist, there is 
one of particular interest, the so-called exotic heavy-light four-quark sector. The 
possible existence of stable $QQ\bar q\bar q$ states has been addressed using different 
approaches since the pioneering work of Ref.~\cite{Ade82}. Exotic heavy-light
four-quark states represent a very interesting exception in the landscape of exotic hadronic physics, 
because there is a broad theoretical consensus about its adequacy to lodge bound states 
for large $M_Q/m_q$ ratios. In particular, there is a long-standing prediction, 
strengthened by several independent studies during the last years, about the existence 
of a deeply bound $bb\bar u\bar d$ isoscalar state with quantum numbers 
$J^P=1^+$~\cite{Fra17,Kar17,Eic17,Bic16,Vij09,Ric18,Luo17,Duc13,Cza18,Ade82}. 
In the charm sector, the decrease of the mass ratio $M_Q/m_q$ might give rise 
just to a shallow bound state with the same quantum numbers~\cite{Jan04,Che16}.

In between $bb\bar q\bar q$ and $cc\bar q\bar q$, one finds the case with two 
distinguishable heavy quarks, $bc\bar q\bar q$, which has not received the same attention 
in the literature. The non-identity of the heavy flavors enlarges the Hilbert 
space and, thus, conclusions cannot be straightforwardly extrapolated from the case
of identical heavy flavors. $QQ'\bar q\bar q$ states have been studied in Refs.~\cite{Sil93,Sem94}
solving the four-body problem by expanding the wave function up to eight quanta in a 
harmonic oscillator basis. Two isoscalar $bc\bar u\bar d$ 
states close to threshold were identified as candidates to 
be bound, the $J^P=0^+$ and $1^+$. Note, however, that systematic expansions 
on the eigenstates of a harmonic oscillator 
is not very efficient to account for short-range correlations and could miss 
binding when it is induced by chromomagnetic effects~\cite{Ric18,Jan04}.  
Ref.~\cite{Kar17} has estimated the mass of the isoscalar $J^P=0^+$ $bc\bar u\bar d$ state 
obtaining a central value 11 MeV below the $\bar B D$ threshold, although it is cautioned 
that the precision of the calculation is not sufficient to determine whether the 
tetraquark is actually above or below the corresponding two-meson threshold. 
Unfortunately, this reference has not analyzed the isoscalar $bc\bar u\bar d$ $J^P=1^+$ state.
The interest on exotic heavy-light tetraquarks with non identical heavy-flavors comes reinforced by
the recent estimation of the production rate of the isoscalar $bc\bar u \bar d$ $J^P=1^+$
state at the LHCb, two orders of magnitude larger than that of the $bb\bar u\bar d$ analogous 
state~\cite{Ali18}.

In this work we adopt generic constituent models to address four-quark systems containing 
distinguishable charm and bottom heavy flavors. We use two different methods to 
look for possible bound states, a variational approach with generalized Gaussians and the 
scattering of two mesons with different heavy flavor content. 
The manuscript is organized as follows. In Sec.~\ref{se:var} we outline the relevant properties of the constituent models 
and the methods considered. In Sec.~\ref{se:Results} we present and discuss the results. Finally,
the main conclusions are summarized in Sec.~\ref{se:concl}.

\section{Solving the $bc\bar q\bar q$ system}
\label{se:var}
For the sake of generality and to judge the independence of the results from the particular model considered, 
two different constituent models widely used in the tetraquark literature are implemented. The first one is the AL1 potential 
by Semay and Silvestre-Brac~\cite{Sem94}. It contains a chromoelectric part made of a Coulomb-plus-linear interaction 
together with a chromomagnetic spin-spin term described by a regularized Breit-Fermi interaction with a
smearing parameter that depends on the reduced mass of the interacting quarks.
The second one is the constituent quark cluster (CQC) model of Ref.~\cite{Vij05}. 
Besides chromoelectric and chromomagnetic terms analogous to the AL1 potential, it considers a chiral potential between
light quarks. The main advantage of these models is that they reasonably describe the heavy meson 
spectra and, thus, the thresholds relevant for each particular set of quantum numbers are correctly
described within the same model.

Two different methods are used to tackle the possible existence of four-quark bound states. 
In the first one we use a variational approach, where the wave function is expanded 
as a linear combination of all allowed vectors in color, spin, flavor and radial subspaces. For the radial part
we make use of generalized Gaussians. The basis dimension quickly escalates with the number of allowed vectors 
and therefore the numerical treatment becomes increasingly challenging although tractable. In the second approach, 
an expansion in terms of all contributing physical meson-meson 
states is considered. Within this scheme the meson-meson interaction
is obtained from the quark-quark potential and then a two-body coupled-channel problem is solved.
The equivalence of the two methods for the two-baryon system was theoretically derived in Ref.~\cite{Har81}.
For the two-meson case it has been mathematically proved in Ref.~\cite{Vin09} and numerically
checked in Refs.~\cite{Vin09,Car11}. 
 
To be a bit more specific, let us note that
four-quark systems present a richer color structure than standard 
baryons or mesons. The color wave function for standard hadrons leads 
to a single vector, but dealing with four-quark hadrons there 
are different vectors driving to a color singlet state out of colorless meson-meson ($\bf 1 \bf 1$) or 
colored two-body ($\bf{8} \bf{8}$, $\bf{\bar 3} \bf{3}$, or $\bf{6} \bf{\bar 6}$) components. 
Note, however, that any colored two-body component can be expanded as an infinite sum of colorless
singlet-singlet states~\cite{Har81}. This has been explicitly
done for $QQ\bar q \bar q$ states in Ref.~\cite{Vin09}.

The lowest lying tetraquark configuration for systems 
with two-heavy flavors presents a separate dynamics for the heavy quarks, in a color $\bf \bar 3$
state, and for the light quarks, bound to a color $\bf 3$ state, to construct
a color singlet~\cite{Eic17} (see the probabilities in Table II of Ref.~\cite{Vin09}
for the isoscalar axial vector $bb\bar u \bar d$ tetraquark). 
This argument has been recently revised in Ref.~\cite{Ric18}, showing 
in Fig. 8 how the probability of the 
$\bf 6 \bf \bar 6$ component in a compact $QQ\bar q\bar q$ tetraquark tends 
to zero for $M_Q \to \infty$.
Therefore, heavy-light compact bound states would be a dominant $\bf{\bar 3} \bf{3}$ 
color state and not a single colorless meson-meson molecule, $\bf 1 \bf 1$.
Such compact states with two-body colored components
can be expanded as the mixture of several physical meson-meson 
channels~\cite{Har81} (see Table II of Ref.~\cite{Vin09})
and, thus, they can be also studied as an involved
coupled-channel problem of physical meson-meson states~\cite{Car11,Ike14}.

Let us summarize in the following subsections the main properties of the two methods
used to look for bound states along this work.

\subsection{Four-quark systems}
\label{4q}
The $bc\bar q\bar q$ four-quark problem has been solved following the variational method outlined in Ref.~\cite{Via09}, 
expanding the radial wave function in terms of generalized Gaussians. The constituent model used is AL1. 
The variational wave function must include all possible flavor-spin-color channels
contributing to a given configuration. Thus, for each channel $s$, the wave function will be the tensor product of
color ($\left|C_n\right>$), spin ($\left|S_m\right>$), flavor ($\left|T_k\right>$), and radial
($\left|R_r\right>$) components,
\begin{equation}
\label{efr}
\left| n\, m\, k\, r\right>=\left|C_n\right>\otimes\left|
S_m\right>\otimes\left|T_k\right>\otimes\left|R_r\right> \, .
\end{equation}
Once the color, spin, and flavor parts are integrated out, the coefficients of the radial wave function are
obtained by solving the system of linear equations,
\begin{equation}
\label{funci1g}
\sum_{s'\,s} \sum_{i} \beta_r^{(i)}
\, [\langle R_{r'}^{(j)}|\,H\,|R_r^{(i)}
\rangle - E\,\langle
R_{r'}^{(j)}|R_r^{(i)}\rangle \delta_{s,s'} ] = 0
\qquad \qquad \forall \, j\, ,
\end{equation}
where the eigenvalues are obtained by a minimization procedure.

Let us discuss briefly the different terms outlined in the wave function of Eq.~(\ref{efr}). 
The flavor part is uniquely determined by the isospin of the light
antiquark pair: $\left|T_1\right>$ for $T=0$ and $\left|T_2\right>$ for $T=1$. 
The spin part of the wave function can be written as
$\left[(\tfrac{1}{2}\tfrac{1}{2})_{S_{12}}(\tfrac{1}{2}\tfrac{1}{2})_{S_{34}}\right]_{S}\equiv|S_{12}S_{34}\rangle$,
where the spin of the two quarks (antiquarks) is coupled to $S_{12}$ ($S_{34}$). 
In Table~\ref{spin} we have summarized the vectors contributing to each total spin state, $S$.
\begin{table}[t]
\caption[Spin basis vectors.]{Spin basis vectors, $|S_{12}S_{34}\rangle$, for the different total spin states, $S$.}
\label{spin}
\begin{center}
\begin{tabular}{|ccc|}
\hline
$S=0$                      & $S=1$                      & $S=2$\\
\hline
$|S_1\rangle=|00\rangle$ & $|S_1\rangle=|10\rangle$ & $|S_1\rangle=|11\rangle$ \\
$|S_2\rangle=|11\rangle$ & $|S_2\rangle=|01\rangle$ & \\
                         & $|S_3\rangle=|11\rangle$ & \\
\hline
\end{tabular}
\end{center}
\end{table}

The most general radial wave function with total orbital angular momentum
$L=0$ is constructed as a linear combination of generalized Gaussians 
depending on a set of variational parameters. 
The usual four--body $H-$like Jacobi coordinates are considered,
\begin{eqnarray}
\label{coo}
\vec{x} &=&\vec{r}_{1}-\vec{r}_{2} \\ \nonumber
\vec{y} &=&\vec{r}_{3}-\vec{r}_{4} \\ \nonumber
\vec{z} &=&\frac{m_{1}\vec{r}_{1}+m_{2}\vec{r}_{2}}{m_{1}+m_{2}}-\frac{m_{3}\vec{r}_{3}+m_{4}\vec{r
}_{4}}{m_{3}+m_{4}}\\ \nonumber
\vec{R} &=&\frac{\sum_i m_{i}\vec{r}_{i}}{\sum_i m_{i}}\nonumber \, ,
\end{eqnarray}
\noindent 
where $1$ and $2$ stand for the quarks and $3$ and $4$ for the antiquarks. 
Thus, we define the function,
\begin{equation}
\label{red1}
g(\alpha_1,\alpha_2,\alpha_3)={\rm Exp}\left(-a^i\vec x^{\,2}-b^i\vec y^{\,2}-c^i\vec z^{\,2}
-\alpha_1d^i\vec x \cdot \vec y-\alpha_2e^i\vec x \cdot \vec z-\alpha_3f^i\vec y \cdot \vec z\right),
\end{equation}
and the vectors
\begin{equation}
\overrightarrow{G^i}=\left(\begin{array}{l} g(+,+,+)\\g(-,+,-)\\g(-,-,+)\\g(+,-,-)\end{array}\right)\, ,
\end{equation}
and
\begin{eqnarray}
\label{red2}
\overrightarrow{\alpha_{SS}}&=&(+,+,+,+) \, , \\ \nonumber
\overrightarrow{\alpha_{SA}}&=&(+,-,+,-) \, , \\ \nonumber
\overrightarrow{\alpha_{AS}}&=&(+,+,-,-) \, , \\ \nonumber
\overrightarrow{\alpha_{AA}}&=&(+,-,-,+) \, ,
\end{eqnarray}
where $S(A)$ stands for symmetric (antisymmetric) under the exchange of quarks $1\leftrightarrow 2$ and antiquarks $3\leftrightarrow 4$.
Then, four different radial wave functions can be constructed depending on their permutation properties: 
\begin{eqnarray}
\label{wave2-1}
(SS)\Rightarrow \left|R_1\right>&=&\sum_{i=1}^{n}\beta_1^{(i)}\left(\overrightarrow{\alpha_{SS}}\cdot \overrightarrow{G^i}\right) \, , \\ \nonumber
(SA)\Rightarrow \left|R_2\right>&=&\sum_{i=1}^{n}\beta_2^{(i)}\left(\overrightarrow{\alpha_{SA}}\cdot \overrightarrow{G^i}\right) \, , \\ \nonumber
(AS)\Rightarrow \left|R_3\right>&=&\sum_{i=1}^{n}\beta_3^{(i)}\left(\overrightarrow{\alpha_{AS}}\cdot \overrightarrow{G^i}\right) \, , \\ \nonumber
(AA)\Rightarrow \left|R_4\right>&=&\sum_{i=1}^{n}\beta_4^{(i)}\left(\overrightarrow{\alpha_{AA}}\cdot \overrightarrow{G^i}\right)\,.
\end{eqnarray}
\noindent The scalar product $\overrightarrow{\alpha_{jk}}\cdot \overrightarrow{G^i}$
generates the appropriate combination of generalized Gaussians to have the specified symmetry $jk$ (see section
2.9 of Ref.~\cite{Via09} for a thorough discussion of the technical details) 
and $n$ is the number of generalized Gaussians required to reach convergence.
The terms mixing Jacobi coordinates, i.e., $\vec{x}\cdotp\vec{y}$, $\vec{x}\cdotp\vec{z}$, and $\vec{y}\cdotp\vec{z}$,
allow for nonzero internal orbital angular momenta, although the total orbital angular momentum is coupled to $L=0$. 
This ensures the positive parity of the states studied.

Finally, regarding the color structure, there are three different ways to couple two quarks and two antiquarks into a colorless state:
\begin{eqnarray}
\label{eq1a}
[(q_1q_2)(\bar q_3\bar q_4)]&\equiv&\{|\bar 3_{12}3_{34}\rangle,|6_{12}\bar 6_{34}\rangle\}\equiv\{|\bar 33\rangle,
|6\bar 6\rangle\} \, ,\\
\label{eq1b}
[(q_1\bar q_3)(q_2\bar q_4)]&\equiv&\{|1_{13}1_{24}\rangle,|8_{13} 8_{24}\rangle\}\equiv\{|11\rangle,|88\rangle\} \, ,\\
\label{eq1c}
[(q_1\bar q_4)(q_2\bar q_3)]&\equiv&\{|1_{14}1_{23}\rangle,|8_{14} 8_{23}\rangle\}\equiv\{|1'1'\rangle,|8'8'\rangle\}\,.
\end{eqnarray}
Each coupling scheme represents a color orthonormal basis where the four-quark problem
can be studied. Only two of these states have well defined
permutation properties: $\ktt$ is antisymmetric under the exchange
of both quarks and antiquarks, and $\kss$ is symmetric. Therefore, the basis~(\ref{eq1a}) is the most
suitable to deal with the Pauli principle. The other two,~(\ref{eq1b}) and~(\ref{eq1c}), 
are hybrid bases containing singlet-singlet (physical) and octet-octet (hidden-color)
vectors, that are required to extract meson-meson physical components 
from the final wave function. In the following, we denote $\left|C_1\right>=\ktt$ and $\left|C_2\right>=\kss$.

The system $bc\bar q\bar q$ contains two identical light antiquarks, 
therefore the Pauli principle has to be applied to this pair. A summary of all vectors allowed 
for the different spin-isospin channels is given in Table~\ref{table2}. 
Further details on the formalism can be obtained from Refs.~\cite{Via09,Vij13}, and references therein.
\begin{table}[t]
\caption{\label{table2} $\left| n\, m\, k\, r\right>$ basis vectors, with the notation of Eq.~(\ref{efr}), 
contributing to each total spin and isospin state, $(S,T)$.}
\begin{center}
\begin{tabular}{|cccccc|}
\hline
$(0,0)$         & $(0,1)$         & $(1,0)$         & $(1,1)$         & $(2,0)$          &$(2,1)$          \\
\hline
$\left|1111\right>$ & $\left|1122\right>$ & $\left|1112\right>$ & $\left|1121\right>$ & $\left|1112\right>$ & $\left|1121\right>$ \\
$\left|1113\right>$ & $\left|1124\right>$ & $\left|1114\right>$ & $\left|1123\right>$ & $\left|1114\right>$ & $\left|1123\right>$ \\
$\left|1212\right>$ & $\left|1221\right>$ & $\left|1211\right>$ & $\left|1122\right>$ & $\left|2111\right>$ & $\left|2122\right>$ \\
$\left|1214\right>$ & $\left|1223\right>$ & $\left|1213\right>$ & $\left|1224\right>$ & $\left|2113\right>$ & $\left|2124\right>$ \\
$\left|2112\right>$ & $\left|2121\right>$ & $\left|1312\right>$ & $\left|1321\right>$ & & \\
$\left|2114\right>$ & $\left|2123\right>$ & $\left|1314\right>$ & $\left|1323\right>$ & & \\
$\left|2211\right>$ & $\left|2222\right>$ & $\left|2111\right>$ & $\left|2122\right>$ & & \\
$\left|2213\right>$ & $\left|2224\right>$ & $\left|2113\right>$ & $\left|2124\right>$ & & \\
                    &                     & $\left|2212\right>$ & $\left|2221\right>$ & & \\
                    &                     & $\left|2214\right>$ & $\left|2223\right>$ & & \\
                    &                     & $\left|2311\right>$ & $\left|2322\right>$ & & \\
                    &                     & $\left|2313\right>$ & $\left|2324\right>$ & & \\
\hline
\end{tabular}
\end{center}
\end{table}

\subsection{Meson-meson systems}
\label{MM}
The $bc\bar q\bar q$ four-quark problem has also been addressed by solving the
Lippmann-Schwinger equation for a two-meson coupled-channel problem.
All allowed meson-meson components made of the lowest 
$S$-wave mesons: $\bar B$, $D$, $\bar B^*$, and $D^*$,
have been considered, see Table~\ref{tabMM}. The number of coupled
channels in the meson-meson approach increases with the number of allowed vectors 
in the four-quark formalism, as seen in Table~\ref{table2}.
\begin{table}[b]
\caption{\label{tabMM} Meson-meson channels contributing to each total spin and isospin state, $(S,T)$.}
\begin{center}
\begin{tabular}{|ccc|}
\hline
$(0,0),(0,1)$            & $(1,0),(1,1)$                           & $(2,0),(2,1)$ \\
\hline
$\bar B D$     & $\bar B^* D$   & $\bar B^* D^*$  \\
$\bar B^* D^*$ & $\bar B D^*$   &                \\
               & $\bar B^* D^*$ &                \\
\hline
\end{tabular}
\end{center}
\end{table}

Thus, we consider a system of two mesons interacting through a potential $V$ that 
has been obtained from the CQC model of Ref.~\cite{Vij05}. If we denote the different meson-meson
systems as channel $A_i$, the Lippmann-Schwinger equation
for the meson-meson scattering becomes,
\begin{eqnarray}
t_{\alpha\beta;JT}^{L_\alpha S_\alpha, L_\beta S_\beta}(p_\alpha,p_\beta;E)& = & 
V_{\alpha\beta;JT}^{L_\alpha S_\alpha, L_\beta S_\beta}(p_\alpha,p_\beta)+
\sum_{\gamma=A_1,A_2, ...}\sum_{L_\gamma=0,2} 
\int_0^\infty p_\gamma^2 dp_\gamma V_{\alpha\gamma;JT}^{L_\alpha S_\alpha, L_\gamma S_\gamma}
(p_\alpha,p_\gamma) \nonumber \\
& \times& \, G_\gamma(E;p_\gamma)
t_{\gamma\beta;JT}^{L_\gamma S_\gamma, L_\beta S_\beta}
(p_\gamma,p_\beta;E) \,\,\,\, , \, \alpha,\beta=A_1,A_2,...\, ,
\label{eq0}
\end{eqnarray}
where $t$ is the two-body scattering amplitude, $J$ and $T$ are the
total angular momentum and isospin of the system,
$L_{\alpha} S_{\alpha}$, $L_{\gamma} S_{\gamma}$, and
$L_{\beta} S_{\beta }$ are the initial, intermediate, and final 
orbital angular momentum and spin, respectively. 
$p_\alpha$ ($p_\beta$) stands for the initial (final) relative
momentum of the two-body system that enters in the Fourier Transform of the potential
in configuration space, and $E$ is the total energy of the two-body system.
$p_\gamma$ is the relative momentum of the intermediate two-body system $\gamma$.
We refer the reader to Ref.~\cite{Rij14} for
a thorough discussion of the technical details
about the solution of the momentum-space Lippmann-Schwinger 
equation.

The propagators $G_\gamma(E;p_\gamma)$ are given by
\begin{equation}
G_\gamma(E;p_\gamma)=\frac{2 \mu_\gamma}{k^2_\gamma-p^2_\gamma + i \epsilon} \, ,
\end{equation}
with
\begin{equation}
E=\frac{k^2_\gamma}{2 \mu_\gamma} \, ,
\end{equation}
where $\mu_\gamma$ is the reduced mass of the two-body system $\gamma$.
For bound-state problems $E < 0$ so that the singularity of the propagator
is never touched and we can forget the $i\epsilon$ in the denominator.
If we make the change of variables
\begin{equation}
p_\gamma = b\frac{1+x_\gamma}{1-x_\gamma},
\label{eq2}
\end{equation}
where $b$ is a scale parameter, and the same for $p_\alpha$ and $p_\beta$, we can
write Eq.~(\ref{eq0}) as,
\begin{eqnarray}
t_{\alpha\beta;JT}^{L_\alpha S_\alpha, L_\beta S_\beta}(x_\alpha,x_\beta;E)& = & 
V_{\alpha\beta;JT}^{L_\alpha S_\alpha, L_\beta S_\beta}(x_\alpha,x_\beta)+
\sum_{\gamma=A_1,A_2,...}\sum_{L_\gamma=0,2} 
\int_{-1}^1 b^2\left(\frac{1+x_\gamma}{1-x_\gamma}\right)^2 \,\, \frac{2b}{(1-x_\gamma)^2}
dx_\gamma \nonumber \\
&\times & V_{\alpha\gamma;JT}^{L_\alpha S_\alpha, L_\gamma S_\gamma}
(x_\alpha,x_\gamma) \, G_\gamma(E;p_\gamma) \,
t_{\gamma\beta;JT}^{L_\gamma S_\gamma, L_\beta S_\beta}
(x_\gamma,x_\beta;E) \, .
\label{eq3}
\end{eqnarray}
We solve this equation by replacing the integral from $-1$ to $1$ by a
Gauss-Legendre quadrature which results in the set of
linear equations,
\begin{equation}
\sum_{\gamma=A_1,A_2,...}\sum_{L_\gamma=0,2}\sum_{m=1}^N
M_{\alpha\gamma;JT}^{n L_\alpha S_\alpha, m L_\gamma S_\gamma}(E) \, 
t_{\gamma\beta;JT}^{L_\gamma S_\gamma, L_\beta S_\beta}(x_m,x_k;E) =  
V_{\alpha\beta;JT}^{L_\alpha S_\alpha, L_\beta S_\beta}(x_n,x_k) \, ,
\label{eq4}
\end{equation}
with
\begin{eqnarray}
M_{\alpha\gamma;JT}^{n L_\alpha S_\alpha, m L_\gamma S_\gamma}(E)
& = & \delta_{nm}\delta_{L_\alpha L_\gamma} \delta_{S_\alpha S_\gamma}
- w_m b^2\left(\frac{1+x_m}{1-x_m}\right)^2{\frac{2b}{(1-x_m)^2}} \nonumber \\
& \times & V_{\alpha\gamma;JT}^{L_\alpha S_\alpha, L_\gamma S_\gamma}(x_n,x_m) 
\, G_\gamma(E;{p_\gamma}_m),
\label{eq5}
\end{eqnarray}
and where $w_m$ and $x_m$ are the weights and abscissas of the Gauss-Legendre
quadrature while ${p_\gamma}_m$ is obtained by putting
$x_\gamma=x_m$ in Eq.~(\ref{eq2}).
If a bound state exists at an energy $E_B$, the determinant of the matrix
$M_{\alpha\gamma;JT}^{n L_\alpha S_\alpha, m L_\gamma S_\gamma}(E_B)$ 
vanishes, i.e., $\left|M_{\alpha\beta;JT}(E_B)\right|=0$.

\subsection{Thresholds}
One of the most important aspects for stability studies on tetraquark spectroscopy, often overlooked 
in the literature, is the determination of the two-meson break-up 
thresholds using the same interacting model, hypothesis and approximations considered 
for the four-quark or two-meson study. Due to the presence of 
heavy quarks of different flavors, no antisymmetry restrictions apply to the final 
two-meson states, therefore all possible isospin values,
$T=0$ and $1$, are allowed for each spin state. The structure and energy of the
different $S$-wave thresholds contributing to each spin state are given in Table~\ref{table_t}.

\section{Results}
\label{se:Results}
\begin{figure*}[t]
\vspace*{-1.5cm}
\resizebox{8.cm}{11.cm}{\includegraphics{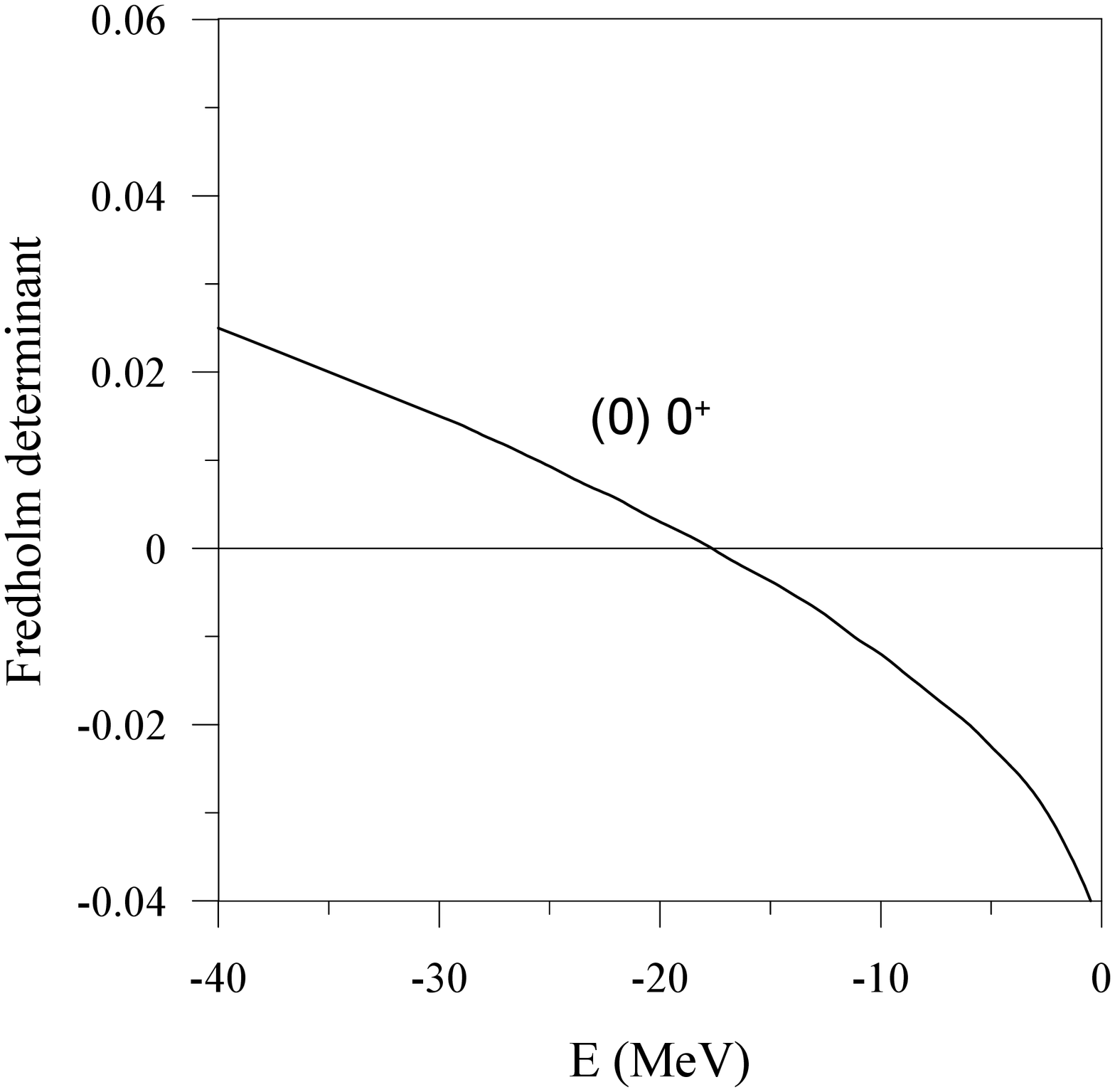}}
\resizebox{8.cm}{11.cm}{\includegraphics{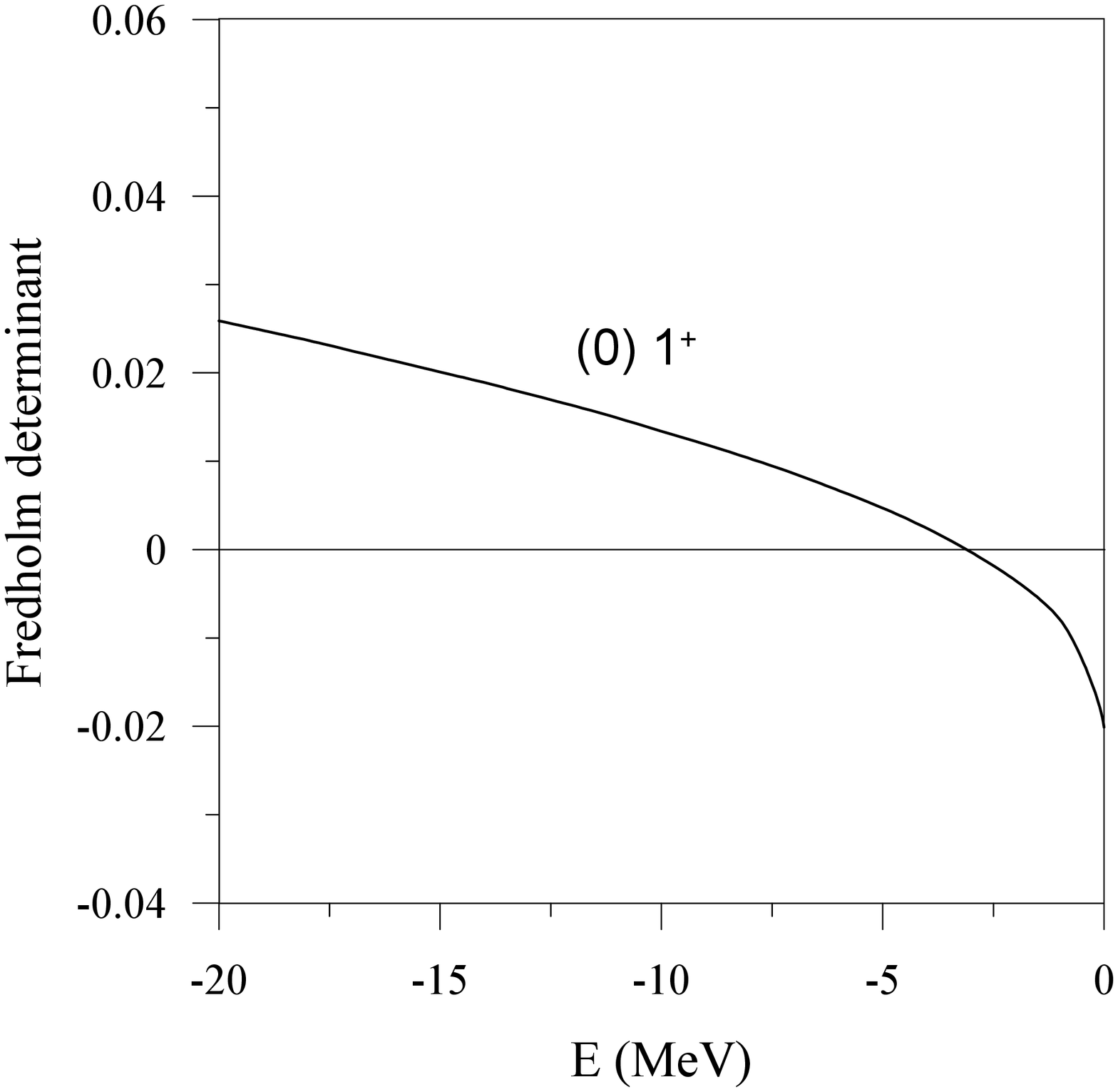}}\vspace*{-3.5cm}
\resizebox{8.cm}{11.cm}{\includegraphics{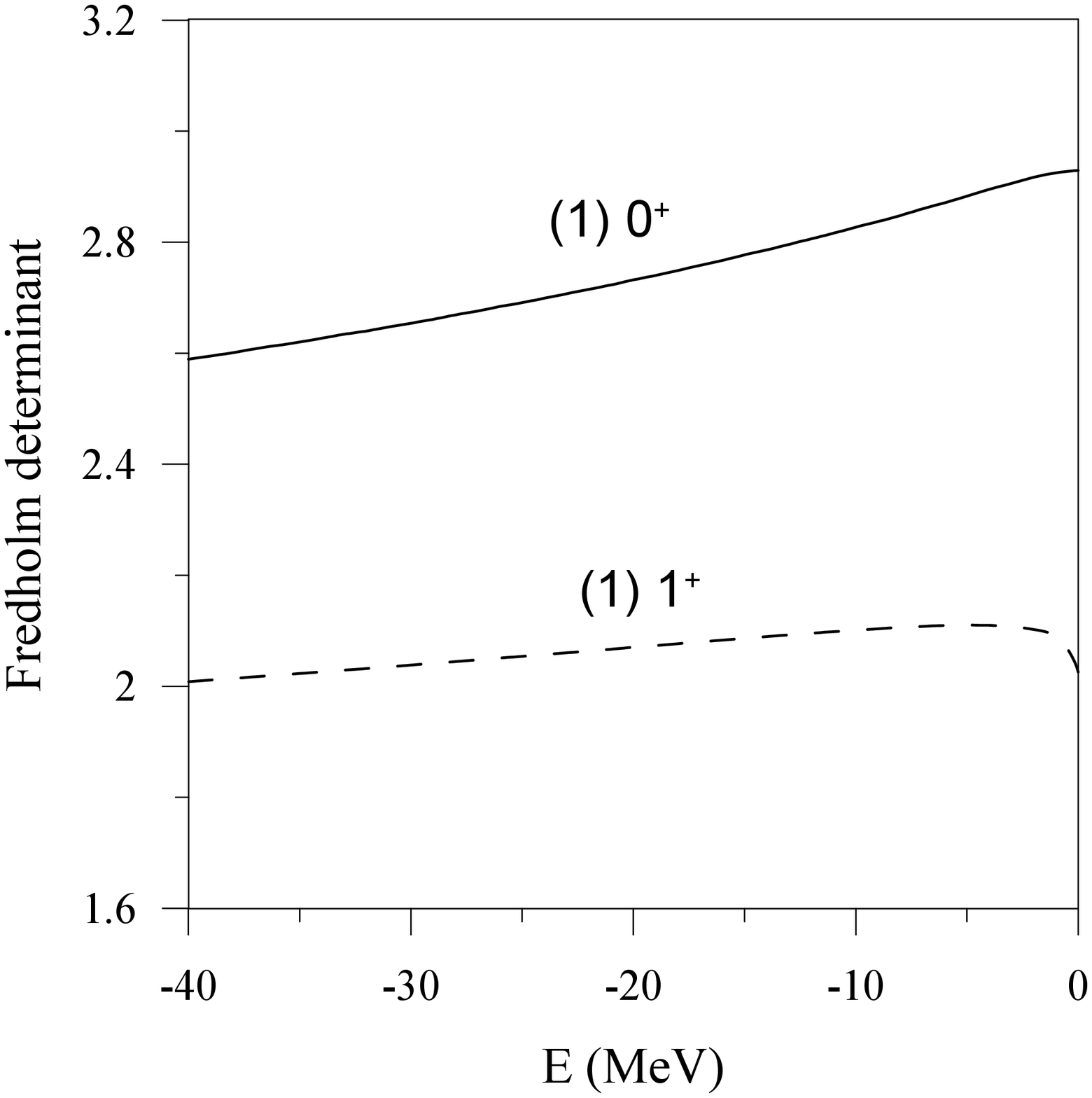}}
\resizebox{8.cm}{11.cm}{\includegraphics{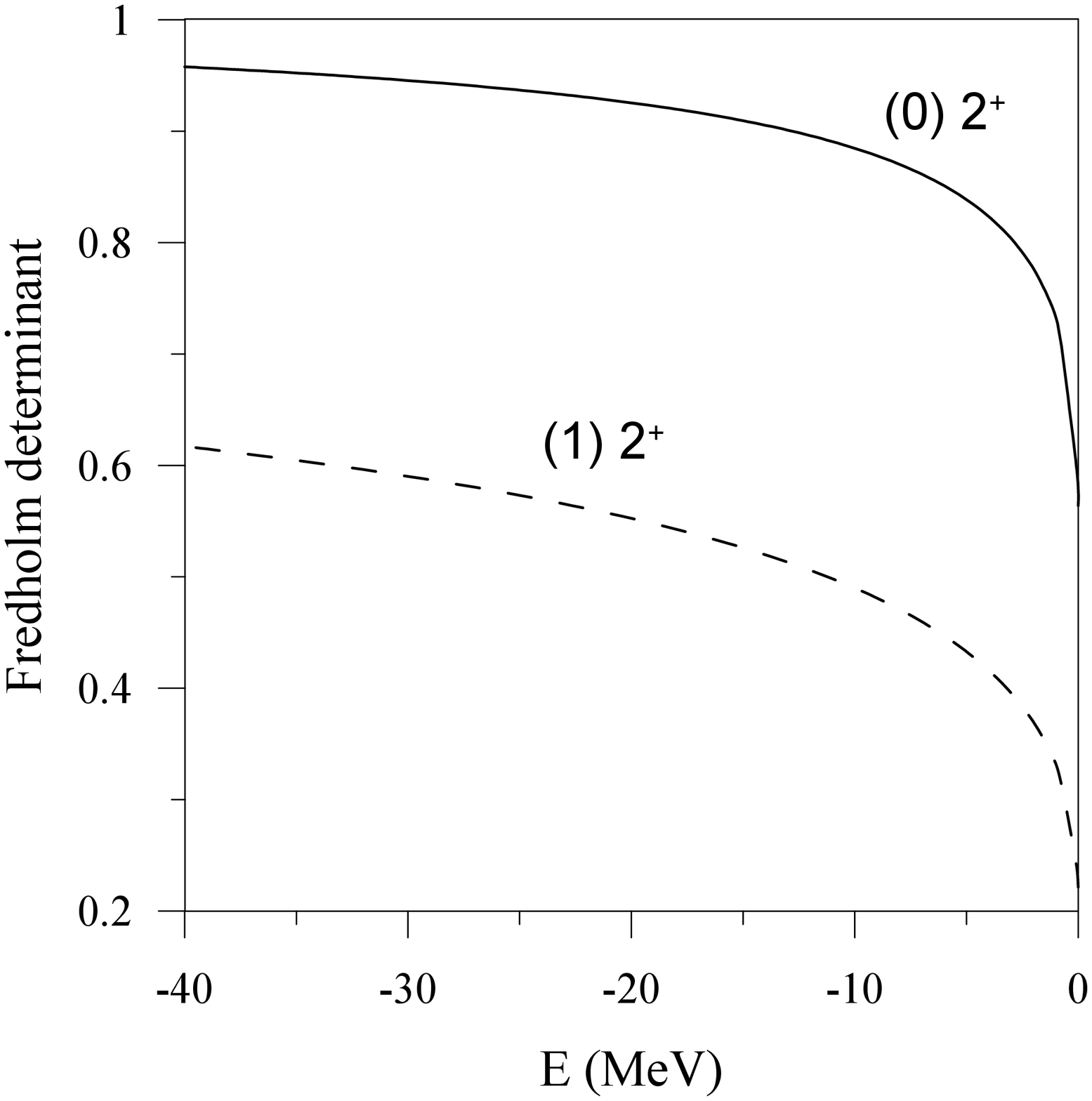}}
\vspace*{-2.0cm}
\caption{Fredholm determinant of the different $(T)J^P$ $bc\bar q\bar q$ channels.}
\label{fig:F1}
\end{figure*}

The results obtained following the procedure outlined in subsection~\ref{4q} are 
given in Table~\ref{Table_res_4q}. In all cases the results have been converged 
till the energy difference between $n-1$ and $n$ generalized Gaussians is less than 1 MeV.
Using six generalized Gaussians the energy of all states is fully converged.
\begin{table*}[t]
\begin{minipage}{\textwidth}
\caption{$S$-wave thresholds for each total spin, $S$, state$^{ 1}$.
Energies are in MeV.}
\begin{center}
\begin{tabular}{|ccc|}
\hline     
$S=0$		 & $S=1$ 		& $S=2$ \\
\hline
$\bar B D$ (7155)     &   			              & 	                  \\
			               & $\bar B^* D$ (7212)   & 	                  \\
			               & $\bar B D^*$ (7309)   & 	                  \\
$\bar B^* D^*$ (7366) & $\bar B^* D^*$ (7366) & $\bar B^* D^*$ (7366) \\
\hline
\end{tabular}
\end{center}
\label{table_t}
\end{minipage}
\end{table*}
\footnotetext[1]{For the sake of simplicity only the energies 
corresponding to the AL1 constituent model are shown, those of the CQC being 
rather similar~\cite{Vij05}.}
\begin{table}[b]
\caption{Four-quark energy, $E_{4q}$, probability of the different color channels, $P_{C_1}$ and $P_{C_2}$, energy of
the lowest threshold, $E_{Th}$, and binding energy, $B=E_{4q}-E_{Th}$, of the different $bc\bar q\bar q$ spin-isospin 
states, $(S,T)$, obtained using the AL1 model. Energies are in MeV.}\label{Table_res_4q}
\begin{center}
\begin{tabular}{|c|c|cc|c|c|}     
    \hline
     $(S,T)$      & $E_{4q}$      & $P_{C_1}$       & $P_{C_2}$       & $E_{Th}$      & $B$ \\
     \hline
     $(0,0)$      & 7132          & 0.49            &   0.51          & 7155          & $-$23 \\
     $(0,1)$      & 7194          & $\frac{1}{3}$   & $\frac{2}{3}$   & 7155          & $+$39 \\     
     $(1,0)$      & 7189          & 0.61            &   0.39          & 7212          & $-$23 \\
     $(1,1)$      & 7245          & $\frac{1}{3}$   & $\frac{2}{3}$   & 7212          & $+$33 \\
     $(2,0)$      & 7363          & 0.26            &   0.74          & 7366          & $-$3 \\
     $(2,1)$      & 7383          & $\frac{1}{3}$   & $\frac{2}{3}$   & 7366          & $+$17 \\
\hline 
\end{tabular}
\end{center}
\end{table}
The results obtained with the meson-meson formalism of subsection~\ref{MM} are shown in 
Fig.~\ref{fig:F1}, where we have plotted the Fredholm determinant as a function of the 
energy, being $E=0$ the mass of the lowest threshold allowed for the corresponding 
channel.

The overall conclusion that can be drawn from Table~\ref{Table_res_4q} and Fig.~\ref{fig:F1} is 
the existence of isoscalar bound states with $J^P=0^+$ and 
$J^P=1^+$, independently of the constituent model considered. Note that although
both constituent models correctly reproduce the meson masses entering the thresholds
which guarantees the similarity of the meson wave functions, there are particularities 
that coherently explain the small differences between them. The CQC model of 
Ref.~\cite{Vij05} contains boson exchanges between the light quarks and thus a weaker
one-gluon exchange chromomagnetic interaction. The slightly smaller bindings 
derived from the CQC model do also give rise to the unbound nature of the 
isoscalar $J^P=2^+$ channel, barely bound with the AL1 model. In both approaches, isovector states are found to be unbound
precluding the existence of charged counterparts. The $J^P=0^+$ state would be strong and electromagnetically stable,
while the $J^P=1^+$ would decay electromagnetically to $\bar B D \gamma$. 
It is worth noting that the stability of the isoscalar
$J^P=0^+$ $bc\bar u\bar d$ state has recently been suggested in Ref.~\cite{Kar17} with a central
value for its mass 11 MeV below the $\bar B D$ threshold, 
although, as mentioned in the introduction, it is cautioned 
that the precision of the calculation is not sufficient to determine whether the  $bc \bar u \bar d$ 
tetraquark is actually above or below the corresponding two-meson threshold. Anyhow, it could
manifest itself as a narrow resonance just at threshold. Unfortunately, this reference
has not analyzed the $J^P=1^+$ state.
The production rate of the isoscalar $bc\bar u \bar d$ $J^P=1^+$
state at the LHCb has been recently estimated in Ref.~\cite{Ali18}.
They have obtained a cross section two orders of magnitude larger 
than that of the production of the $bb\bar u\bar d$ analogous state.
Thus, exotic heavy-light tetraquarks with 
non identical heavy-flavors have an excellent discovery potential at the LHCb.

At first glance, the $bc\bar q\bar q$ system may look deceptively similar to the $cc\bar q\bar q$ and 
$bb\bar q\bar q$ ones. However, the existence of two distinguishable heavy quarks
is a major difference. This makes possible that a large number of 
basis vectors contributes to a particular set of quantum numbers, some of which are 
forbidden in a system with identical heavy flavors. As we discuss below they are relevant to understand
the dynamics of the $bc\bar q\bar q$ bound states. 
\begin{figure}[t]
\centering
\vspace*{-4cm}
\includegraphics[width=0.8\columnwidth]{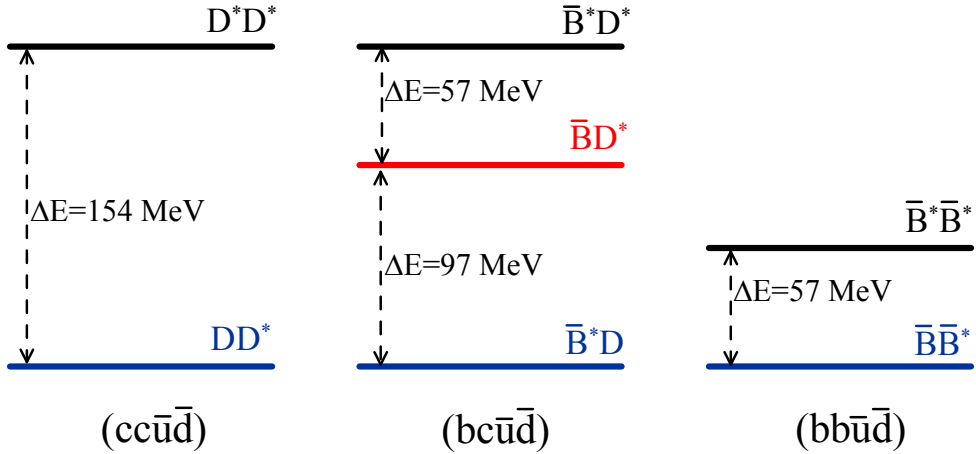}
\vspace*{-8.5cm}
\caption{Two-meson thresholds for the isoscalar $J^P=1^+$ $cc\bar u\bar d$, $bc\bar u\bar d$, and $bb\bar u\bar d$ states.}
\label{fig:F5}
\end{figure}

Let us analyze how the dynamics of thresholds, see Fig.~\ref{fig:F5}, consequence of the larger Hilbert space,
helps in understanding the results obtained for the $bc\bar q\bar q$ system~\cite{Vij14}.
For this purpose, and without loss of generality, we restrict ourselves to the isoscalar axial vector $J^P=1^+$
bound state, existing both in the sector with identical and non-identical heavy flavors. 
We consider the AL1 model, where the $bb\bar u\bar d$ state 
is bound by about 150 MeV~\cite{Ric18} while the $cc\bar u\bar d$ state is bound by about
3 MeV~\cite{Jan04}. 
The chromomagnetic interaction, suppressed by $M_Q$, is stronger in the charm sector
than in the bottom one, what generates larger matrix elements between color-spin vectors
of the pseudoscalar-vector and vector-vector two-meson thresholds.
However, the mass difference between the allowed thresholds increases 
from 57 MeV in the bottom sector to 154 in the charm one, what weakens
the coupling between different color-spin vectors
of the pseudoscalar-vector and vector-vector two-meson thresholds.
Taking into account that the single channel problem of $D D^*$
or $\bar B \bar B^*$ mesons does not present a bound state~\cite{Via09,Car11},
the weaker coupling between $ D D^*$ and $ D^* D^*$ 
than between $\bar B\bar B^*$ and $\bar B^*\bar B^*$, drives to
a reduction of the binding energy from 150 MeV to 3 MeV. If we now consider the isoscalar $bc\bar u\bar d$ $J^P=1^+$
state, the mass difference between the $\bar B^* D$ and $\bar B^* D^*$ thresholds is the same as in the charm case, but
the chromomagnetic interaction involving the bottom quark is weakened by a factor $m_b/m_c \sim 3$. Then, one would 
expect to get a smaller binding energy than in the charm sector. However, the results shown in Table~\ref{Table_res_4q}
exhibit a different trend, with a larger binding energy. The nice
feature of the $bc\bar q\bar q$ state is that it contains distinguishable heavy quarks and thus a new 
threshold (a larger Hilbert space in the language of four-quark states) appears in the $J^P=1^+$ state, 
the $\bar B D^*$, in between $\bar B^* D$ and $\bar B^* D^*$.
Although the $\bar B^* D$ and $\bar B D^*$ systems cannot couple directly, nevertheless, they are coupled 
through the higher $\bar B^* D^*$ state, i.e. $\bar B^* D\leftrightarrow \bar B^* D^* \leftrightarrow \bar B D^*$.
Being the mass difference between $\bar B^* D$ and $\bar B D^*$ smaller than between $D D^*$ and $D^* D^*$
the mixing is reinforced as compared to the charm case, driving to a binding energy larger than in the 
charm sector. The dynamics of thresholds to enhance or diminish coupled-channel effects has been
illustrated at lenght in the literature~\cite{Vij14,Car16,Lut05,Bar15}, although to the best of our knowledge this is the first
example where the presence of an additional intermediate threshold induced by the non-identity of the heavy quarks
helps in increasing the binding.
\begin{figure}[t]
\vspace*{-2.5cm}
\hspace*{-0.8cm}\resizebox{8.5cm}{11.8cm}{\includegraphics{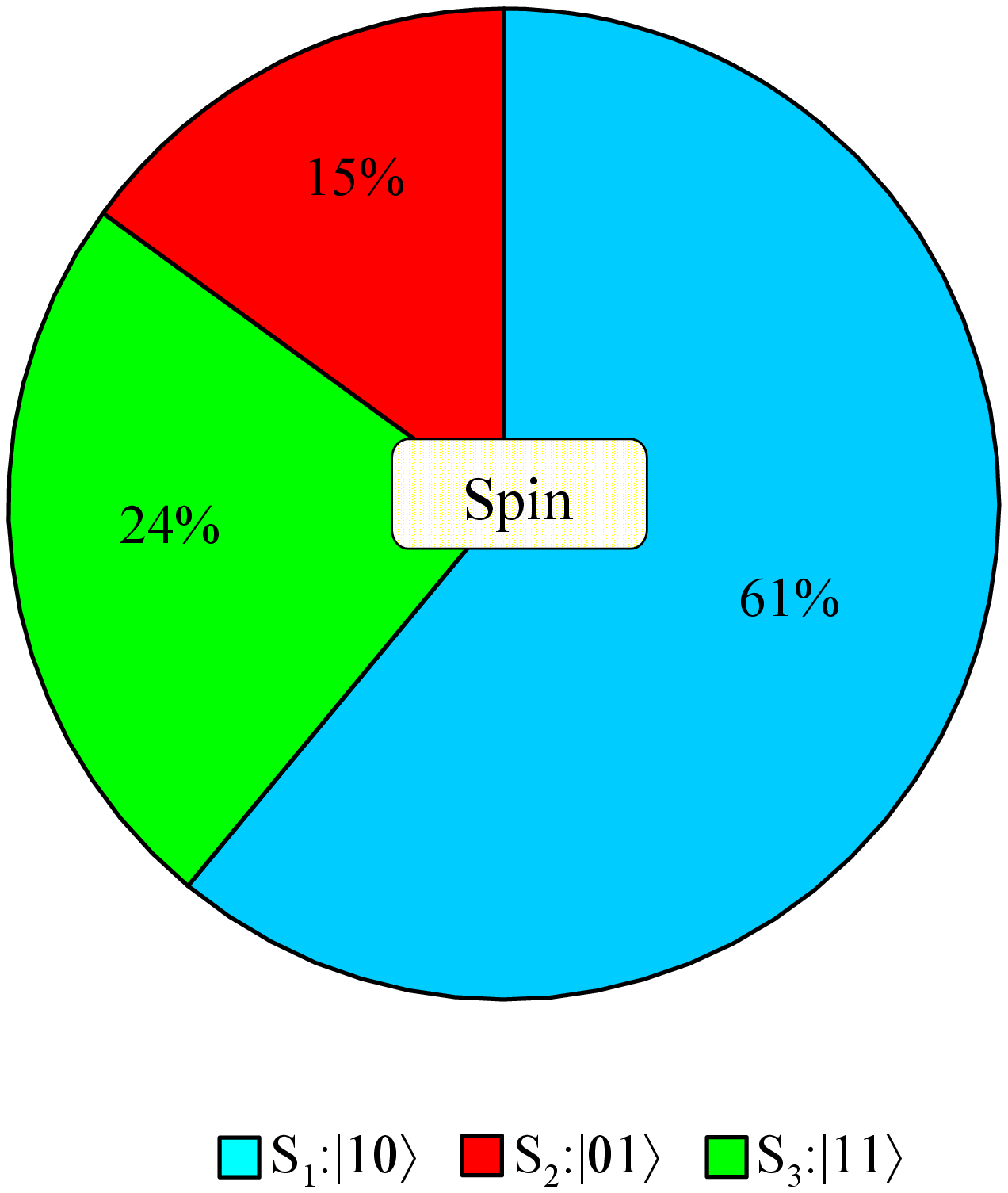}}
\resizebox{8.5cm}{11.8cm}{\includegraphics{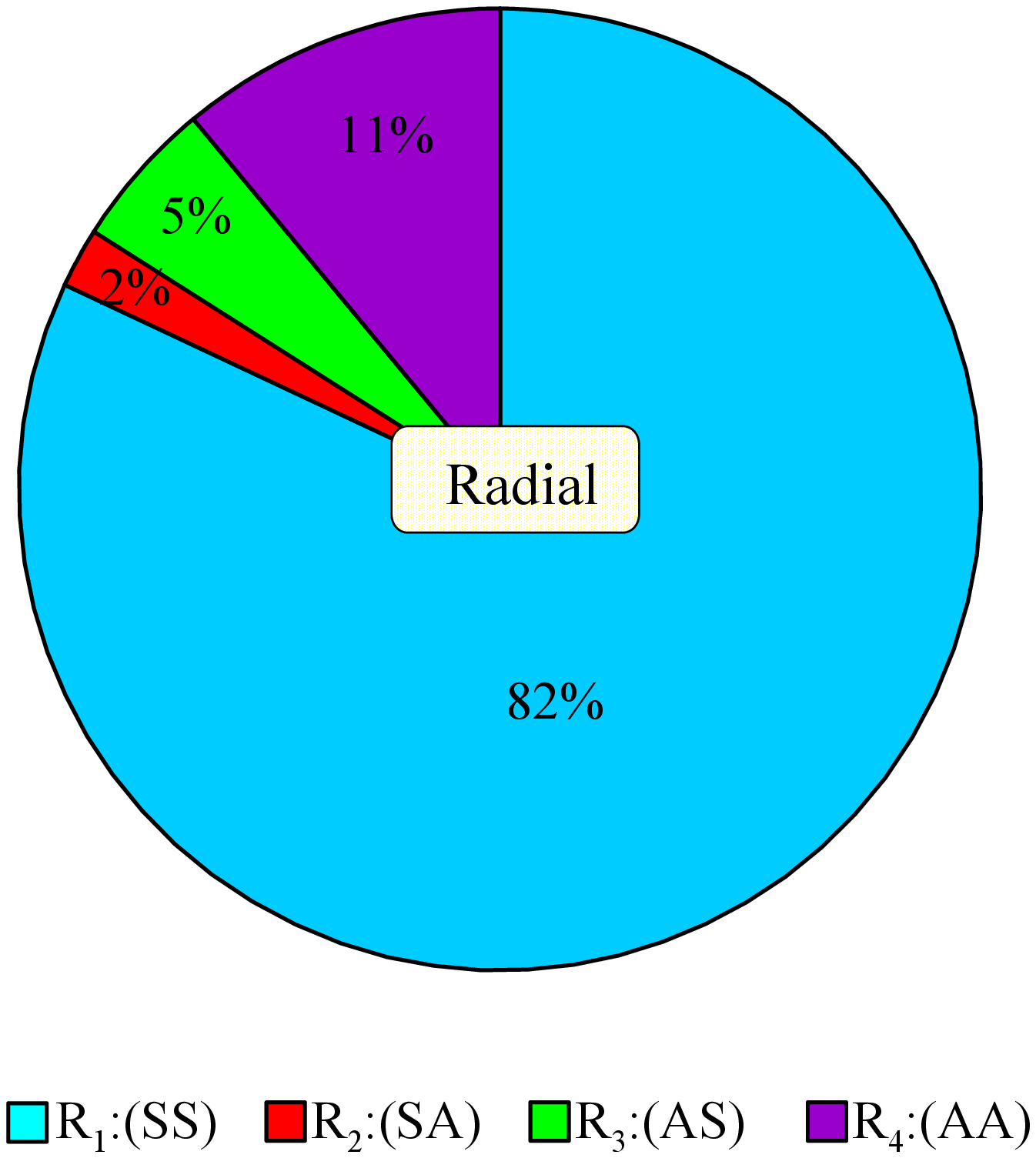}}\vspace*{-4.cm}
\hspace*{-0.8cm}\resizebox{8.5cm}{11.8cm}{\includegraphics{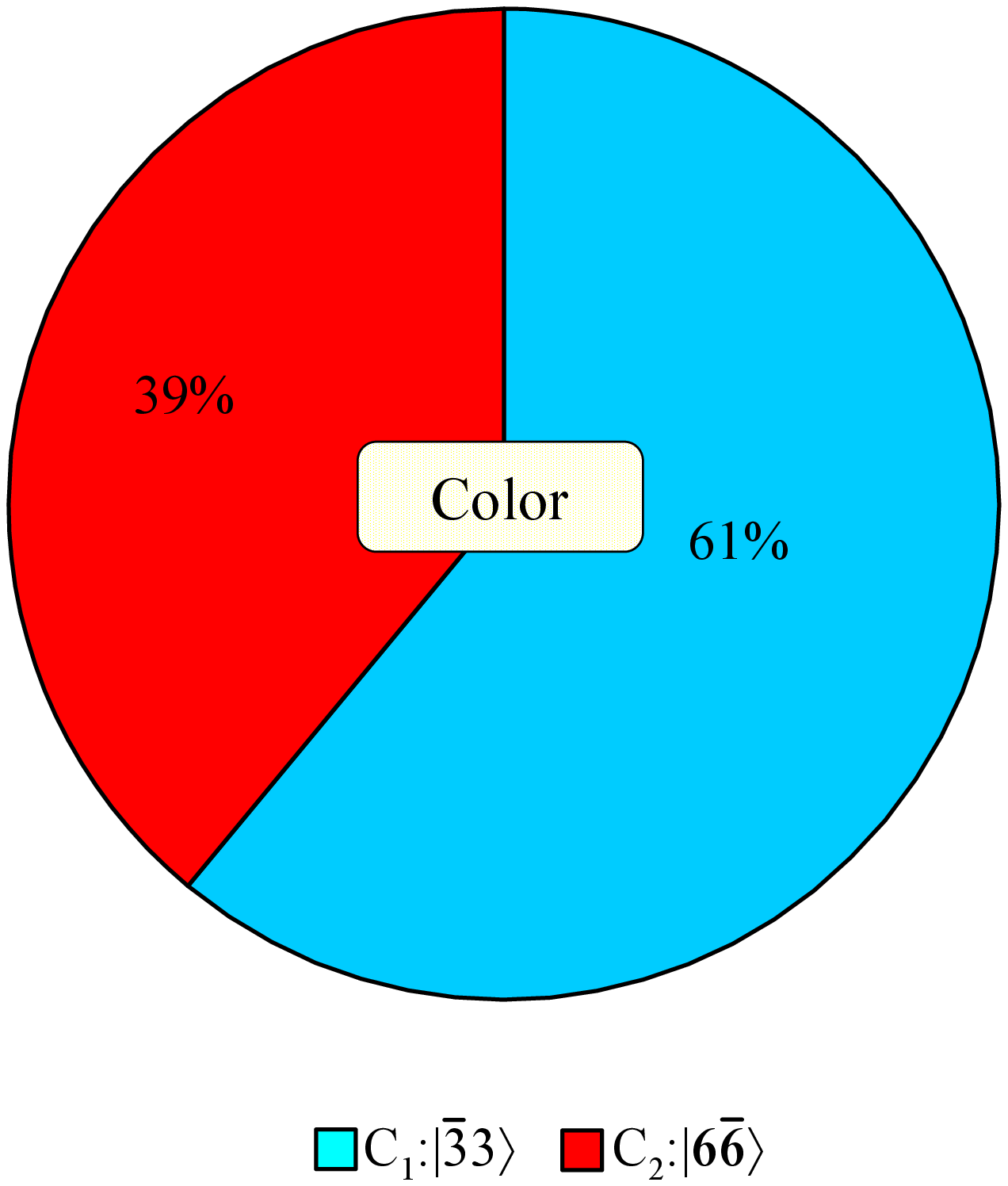}}
\resizebox{8.5cm}{11.8cm}{\includegraphics{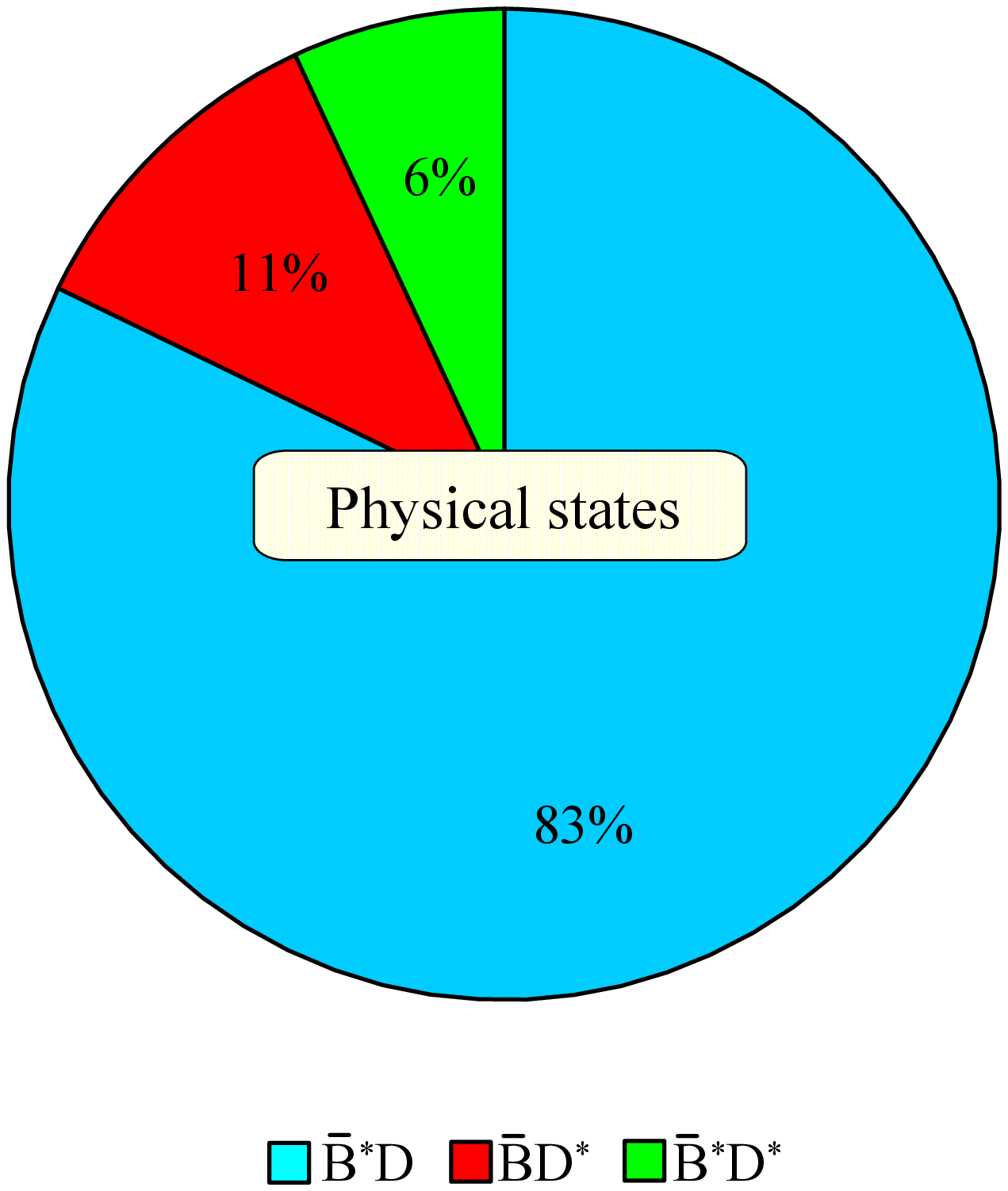}}
\vspace*{-2.0cm}
\caption{Detailed structure of the isoscalar $bc\bar u\bar d$ $J^P=1^+$ wave function.
The first three panels show the probability of the different spin, radial and color 
vectors. The last panel shows the decomposition of the wave function in terms of the 
singlet-singlet color vectors of bases~(\ref{eq1b}) and~(\ref{eq1c}).}
\label{fig:F6}
\end{figure}

Thus, the connection between the two proposed methodologies is amazing and it 
can be analytically derived through the formalism developed in Ref.~\cite{Vin09}. It allows 
to extract the probabilities of meson-meson physical channels out of a four-quark wave function 
expressed as a linear combination of color-spin-flavor-radial vectors. We show in Fig.~\ref{fig:F6} a 
summary of the color, spin, radial, and meson-meson 
component probabilities for the isoscalar $J^P=1^+$ $bc\bar u\bar d$ bound state. 
It is worth noting the 11\% probability of the $\bar B D^*$ component, induced by the indirect coupling
to the lowest $\bar B^* D$ state through the highest $\bar B^* D^*$ component.
As has been recently discussed~\cite{Ric18}, these results present sound evidence
about the importance of including a complete basis, i.e., not discarding any set of 
components a priori. In addition, the importance of a complete radial wave function considering 
terms mixing Jacobi coordinates, thus able to accommodate the antisymmetric terms reported 
in Fig.~\ref{fig:F6}, becomes apparent. 
Unless it is done that way, one is in front of approximations driving to unchecked results~\cite{Ric18}. In particular, 
removing the $6\bar 6$ color components or antisymmetric terms in the wave function, leads to 
unbound states for all quantum numbers.

\section{Outlook}
\label{se:concl}
Heavy-light four-quark states containing a pair of identical $b$ or $c$ heavy 
quarks have been widely discussed in the literature for the last 40 years. However, this has 
not been the case when heavy quarks of different flavor are considered. Thus, in this work 
the possible existence of  $bc\bar q\bar q$ bound states has been addressed. 

Independently of the constituent model used, isoscalar
states are found to be attractive, while isovector states are repulsive, what
precludes the existence of exotic charged heavy-light four-quark states with
distinguishable heavy flavors. The isoscalar $J^P=1^+$ state, holding a bound state for the case 
of identical bottom quarks, is found to be bound in the $bc\bar u\bar d$ case. Besides,
the isoscalar $J^P=0^+$ state, forbidden in $S$-waves for identical heavy flavors, is also 
found to be bound. These two states are bound independently of the constituent model used. 
While the $J^P=0^+$ state would be strong and electromagnetic-interaction stable, the $J^P=1^+$
would decay electromagnetically to $\bar B D \gamma$.
Recent estimations of the production rate of double heavy tetraquarks at the LHCb conclude
the enhancement of the production of non-identical heavy flavors $bc$ compared to the identical
bottom case by two orders of magnitude. In particular, with the LHCb integrated 
luminosity of 50 fb$^{-1}$, to be reached 
in Runs $1-4$,  well over $10^9$ $bc\bar u \bar d$ events will be produced~\cite{Ali18}.

In spite of the supposed similarity with the case of identical heavy flavors, the dynamics
is richer and the interplay among different thresholds drive to unexpected results, as it is
the large binding of the isoscalar axial vector state and the existence of a strong and 
electromagnetic-interaction stable isoscalar scalar state. It is hoped that
the relevance of the present predictions for the understanding of basic properties of low energy
QCD and the current capability of existing experiments, like the LHCb, to detect
these exotic structures, would encourage experimentalists 
to investigate heavy-light four-quark systems also for the case of non-identical
heavy flavors.

\vspace*{1cm}
\noindent
{\bf {\small NOTE ADDED IN PROOF}}\\ 

While this paper was in review, other 
independent calculations made in different frameworks 
arrived to similar conclusions. Among them, it is important to emphasize 
that the lattice QCD results of Ref.~\cite{Fra18}
find evidence for the existence of a strong-interaction-stable $(T)J^P=(0)1^+D$ 
$ud\bar c \bar b$ four-quark state with a mass in the range of 15 to 61 MeV below 
the $\bar DB^*$ threshold. \\

\hrule

\section{Acknowledgments}
This work has been funded by Ministerio de Econom\'\i a, Industria y Competitividad
and EU FEDER under Contracts No. FPA2016-77177.

\end{document}